# The pragmatics of clone detection and elimination

Simon Thompson[a], Huiqing Li[a], and Andreas Schumacher[b]

a    School of Computing, University of Kent, UK
b    Ericsson AB

**Abstract**    The occurrence of similar code, or 'code clones', can make program code difficult to read, modify and maintain. This paper describes industrial case studies of clone detection and elimination, and were were performed in collaboration with engineers from Ericsson AB using the refactoring and clone detection tool Wrangler for Erlang.

We use the studies to illustrate the complex set of decisions that have to be taken when performing clone elimination in practice; we also discuss how the studies have informed the design of the tool. However, the conclusions we draw are largely language-independent, and set out the pragmatics of clone detection and elimination in real-world projects as well as design principles for clone detection decision-support tools.

*Context.* The context of this work is the fact that a software tool is designed to be used; the success of such a tool therefore depends on its suitability and usability in practice.

The work proceeds by observing the use of a tool in particular case studies in detail, through a "participant observer" approach, and drawing qualitative conclusions from these studies, rather than collecting and analysing quantitative data from a larger set of applications. Our conclusions help not only programmers but also the designers of software tools.

*Inquiry.* Data collected in this way make two kinds of contribution. First, they provide the basis for deriving a set of questions that typically need to be answered by engineers in the process of removing clones from an application, and a set of heuristics that can be used to help answer these questions. Secondly, they provide feedback on existing features of software tools, as well as suggesting new features to be added to the tools.

*Approach.* The work was undertaken by the tool designers and engineers from Ericsson AB, working together on clone elimination for code from the company.

*Knowledge.* The work led to a number of conclusions, at different levels of generality.

At the top level, there is overwhelming evidence that the process of clone elimination cannot be entirely automated, and needs to include the input of engineers familiar with the domain in question.

Furthermore, there is strong evidence that the automated tools are sensitive to a set of parameters, which will differ for different applications and programming styles, and that individual clones can be over- and under-identified: again, involving those with knowledge of the code and the domain is key to successful application.

*Grounding.* The work is grounded in "participant observation" by the tool builders, who made detailed logs of the processes undertaken by the group.

*Importance.* The work gives guidelines that assist an engineer in using clone detection and elimination in practice, as well as helping a tool developer to shape their tool building. Although the work was in the context of a particular tool and programming language, the authors would argue that the high-level knowledge gained applies equally well to other notions of clone, as well as other tools and programming languages.

# The Art, Science, and Engineering of Programming

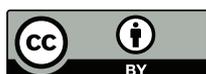

**The pragmatics of clone detection and elimination**

## 1 Introduction

In this paper we explore a case study of clone detection and elimination in practice. The subject of the case study is test code written in the Erlang programming language, and provided by Ericsson.[1] We follow this up with a discussion of further interactions with Ericsson developers who use Erlang. The paper draws a number of conclusions about the process, from general conclusions that the process must be driven by a domain expert, and that it cannot be automated in a "push-button" way to particular observations about the order in which clones should be identified, and some of the pitfalls of the clone identification process.

The case study uses Wrangler [37], a tool that supports interactive refactoring of Erlang programs. Wrangler itself is implemented in Erlang, and supports a variety of refactorings, as well as a set of "code smell" inspection functionalities, and facilities to detect and eliminate code clones. It is integrated with Emacs and Eclipse, through the ErlIDE plugin [8], as well as being available from the command line.

Why is test code particularly prone to clone proliferation? One reason is that many people touch the code: a first few tests are written, and then others author more tests. The quickest way to write these is to copy, paste and modify an existing test, even if this is not the best way to structure the code, it can be done with a minimal understanding of the code. This observation applies equally well to long-standing projects, particularly with a large element of legacy code.

What comes out very clearly from the case study is the fact that refactoring or clone detection cannot be completely automated. In a preliminary experiment by one of the Wrangler developers (Thompson) the code was reduced by some 20% using a "slash and burn" approach, simply eliminating clones one by one. The result of this was – unsurprisingly – completely unreadable. It is only with the collaboration of project engineers, Lindberg and Schumacher, that we were able to identify which clone candidates should be removed, how they could be named and parameterised and so forth. Moreover it requires domain insight to decide on which clones might not be removed. These questions are discussed at length here.

The case study examines the use of a particular definition of "clone", a particular detection mechanism for a particular programming language, as implemented in a particular tool. We would nevertheless argue that this is representative: the type of clone identified is strictly stronger than Type II (and in some ways stronger indeed than Type IV); the language of choice is Erlang, with a focus on sequences of expressions, and so in both these aspects it is comparable with other approaches. We also note that the observations that we make are orthogonal to the particular nature of the algorithm, language and tool, and because of this they are of equal value within other contexts.

The remainder of the paper is organised as follows. Section 2 surveys related work on clone detection and removal, and Section 3 introduces the Erlang programming language and the Wrangler tool. Section 4 describes the support that Wrangler provides for clone detection and removal. Section 5 describes the case study itself, and Section 6 highlights some lessons coming from the case study (cross-referenced with the stages of the case study). Section 7 discusses enhancements made to Wrangler on the basis of further interactions with Ericsson staff, and Section 8 describes a second case study from a different development group within Ericsson. Section 9 shows how clone detection was used in the ProTest project as a mechanism for property extraction and finally, in Section 10 we describe future work and conclude the paper.

---

[1] We are grateful to Ericsson AB for permission to include portions of the code in this paper.

## 2 Clone detection and removal

Duplicated code, or the existence of code clones, is one of the well-known bad code smells when refactoring and software maintenance is concerned. 'Duplicated code', in general, refers to a program fragment that is identical or similar to another, though the exact meaning of 'similar' might vary between different application contexts.

There is an extensive literature on software clones, including a survey of work in the area up to 2007 [32], which *inter alia* covers the standard taxonomy of clones (types I–IV), and there is also a regular *International Workshop on Software Clones* [7, 15].

The most obvious reason for code duplication is the reuse of existing code (by *copy*, *paste* and *modify* for example). Duplicate code introduced for this reason often leads to design problems such as the lack of encapsulation or abstraction. This kind of design problem can be corrected *post hoc* by refactoring out the existing clones at a later stage, or could be avoided entirely by refactoring the existing code to make it reusable without duplication, prior to reusing it [9].

While some code clones might have a sound reason for their existence [6, 17], many clones are considered to be detrimental to the quality of software, because code duplication makes it more likely that bugs will be propagated, as well as increasing the size of the source code and the executable, the compile time, and most importantly the overall cost of maintenance [28, 32].

The existing body of work on software clones places an emphasis on clone identification, analysis and classification – both theoretical and practical [16] – as well as on the evolution and tracking of clones during the lifetime of a project. This literature encompasses theoretical work on different mechanisms of clone identification, as well as case studies that show the presence of code clones in most projects, typically accounting for at least ten percent of the existing code base.

There is less work on clone removal or elimination, and particularly on the pragmatics of what it means, program-by-program. Mechanisms tend to generality, for instance embracing the "near miss" clone [4, 14], using the evolution history of a system [13, 27], adding a degree of paramtricity [19]; or emphasise engineering aspects [36].

The approach of Balazinska and colleagues [3] is to build an automatic tool to replace clones – or at least the clones identified by a particular technique – by uses of the strategy design pattern that allows a class to be configured to behave in one of a number of ways, thereby generalising each particular behaviour. What is notable about this work – but perhaps also its main disadvantage – is that this process is entirely automatic; we will come back to this point when we discuss our case study. An overview of the other work in this area is given in [32]; these tend to vary according to the identification mechanism used, but essentially involve the use of the *Extract Method* refactoring within the OO paradigm.

Other work on the ARIES tool [12] provides analysis together with hints about how clones identified in Java code can be removed, but does not specifically implement those changes. The thesis by Komondoor [18] explores how duplication in source code can be eliminated, but with an emphasis on the identification rather than the elimination of duplicate code; this work builds on earlier studies of duplication in assembly code, but this body of work tends to emphasise compilation technology rather than the production of refactored source code that is recognisable to its author.

The problem is tackled from a different perspective by Göde [10]: he compares the potential candidates for clone elimination that are identified by algorithm to the actual removal of duplicate code that has taken place in a number of projects. In his conclusions he notes

> *Analyzing the removals showed that method extraction was the most frequent refactoring used to eliminate duplication. However, the scope of the refactorings hardly ever matched the scope of detected clones …*
>
> *… in summary, our case study highlights a discrepancy between clones detected by a state-of-the-art tool and the way maintainers approach duplication.*

Tairas and Gray [35] note another discrepancy between theory and practice. In observing a number of open source projects they saw that it was often a sub-clone that was removed, rather than the 'full' clone identified; this is a point that we come back to in Steps 8,9 of Section 5.1, page 16.

Harder and Göde's study [11] about clone management – that is tracking clones once they have been identified – aims to identify the cost of maintaining the clone, and also recommends the involvement of the system developer in decisions about how the clone should be dealt with at appropriate points in its evolution.

Schulze *et al.* [33] observe that it is possible to remove clones in more than one way: specifically they define OOP and AOP approaches, and correlate these with particular kinds of clone. Yu and Ramaswamy [38] study the clones in the Linux distribution: they identify 903 clone sets among the 4635 modules (using a given set of thresholds). Their conclusion is that about 39% of the code clones could be removed by replacing them with calls to a common function, while the others would require techniques beyond the state of the art at the time (2008). They do not investigate the practicalities of removing those clones.

## 3 Erlang and Wrangler

In this section we give an overview of Erlang, sufficient for understanding the examples in the remainder of the paper, as well as giving an introduction to the Wrangler refactoring tool for Erlang.

### 3.1 Erlang

Erlang [1, 5] is a strict, weakly typed functional programming language with built-in support for concurrency, communication, distribution, and fault-tolerance. Erlang is widely used in communication-intensive domains including telecoms, messaging (including WhatsApp Messenger), finance and web applications.

An Erlang system consists of a collection of modules, each of which contains a set of function definitions. A function definition consists of a number of clauses, each of which contains a *head*, involving a sequence of patterns and, separated by '->', a *body*, consisting of a sequence of statements. In this respect they are just like functions in C, or methods in Java. An example of a function definition is given by

```
% Example Erlang function: sum the positive elements in a list.
sum_pos([]) -> 0;
sum_pos([X|Xs]) ->
   V = max(X,0),
   V + sum_pos(Xs).
```

In this example, which sums the positive elements in a list, the pattern in the first clause will match an empty list, whose sum is zero; in the second the pattern matches a non-empty list, with first element (or head) X and remainder (or tail) Xs; for instance, on [2,3,4], X is matched with 2 and Xs with [3,4]. The result of the second clause is the value of the final statement in the body. The symbol % introduces a comment which extends to the end of the line.

The concurrency primitives and others have side-effects, but assignment in Erlang, which is a special case of pattern matching, is single assignment. As shown in the example, Erlang names (which are 'atoms') begin with small letters and variables with capital letters.

Erlang has a single distribution which forms the *de facto* standard. The standard Erlang distribution comes with a number of libraries including the Open Telecom Platform, OTP, as well as more specialised libraries; these include the syntax_tools library that gives an abstract interface to the syntax trees produced by the compiler, as well as the test frameworks EUnit and Common Test. Erlang has an integrated macro language and pre-processor: Erlang macros are conventionally denoted in CAPS and macro calls are preceded by a question mark, thus: ?CAPS.

### 3.2 Wrangler

Wrangler [37] is a refactoring tool for Erlang, designed and implemented in the School of Computing at the University of Kent. It provides a set of structural refactorings, including

- renaming of variables, functions and modules;
- function generalisation;
- function extraction: an identified expression is made the body of a new function definition;

**The pragmatics of clone detection and elimination**

- inlining or 'unfolding' a function application, and its dual, 'folding' an expression into a function application;
- introduce local variables to name an identified expression, and dually, to inline variable definitions,
- introduce new macro definitions and to fold against macro definitions.

In addition, Wrangler provides a set of process-based refactorings, including the introduction of process naming. Wrangler also contains a variety of refactorings which work with Erlang QuickCheck, a property-based testing tool for Erlang.

Wrangler is integrated within Emacs – including XEmacs – and also within Eclipse, as a part of the ErlIDE [8] plugin for Erlang. In implementing Wrangler we have chosen to respect various features of the language and related tools. Wrangler is able to process modules which use *macros*, including the Erlang test frameworks that are in regular use. Wrangler also respects the naming conventions in those test frameworks.

Wrangler provides a portfolio of decision-support tools. The code inspector highlights local 'code smells' and a number of reports highlight issues in the module structure of projects, including circular inclusions and other potential faults [21]. The code clone detection facilities can be used on large multi-module projects to report on code clones and how they can be removed; clone detection can be preformed incrementally on larger code bases, for example as part of a continuous integration approach to software construction [20, 24]. Clone detection is discussed in more detail in the next section.

Wrangler contains a framework that allows users to define for themselves refactorings and code inspection functions that suit their needs [23]. These are defined using a template- and rule-based program transformation and analysis API. Wrangler also supports a domain-specific language that allows users to script composite refactorings, test them and apply them on the fly [25]. User-defined refactorings and scripts are not "second-class citizens": like the existing Wrangler refactorings, user-defined refactorings can be called through the emacs interface and benefit from features such as results preview, layout preservation, selective refactoring, and undo.

## 4 Clone detection and removal with Wrangler

Wrangler provides a 'similar' code detection approach based on the notion of *anti-unification* [29] to detecting code clones in Erlang programs, as well as a mechanism for automatic clone elimination under the user's control.

### 4.1 Foundations

Wrangler provides facilities for finding both 'identical' and 'similar' code. Two pieces of code are said to be *identical* if they are the same when the values of literals and the names of variables are ignored, and the binding structures (of variables to their definitions) are the same. We concentrate on the more general notion of *similarity* here and in the discussion of the case studies.

The *anti-unifier* of two terms denotes their *least-general common abstraction*, and therefore captures the common syntactical structure of the two terms. In general, we say that two expressions (or expression sequences), A and B, are *similar* if there exists a non-trivial least-general common abstraction, C, and two substitutions $\sigma_1$ and $\sigma_2$ which take C to A and B, respectively. We discuss what we mean by 'non-trivial' below.

This clone detection approach is able, for example, to spot that the two expressions ((X+3)+4) and (4+(5-(3*X))) are similar as they are both instances of the expression (Y+Z),[2] and so can both be replaced by calls to the function

```
add(Y,Z) -> Y+Z.
```

Note that the example here is not chosen as a potential candidate for clone elimination, but instead as a suitable example when explaining the notions of substitution and anti-unification.

How does our clone definition fit with the traditional classification of clones into Types I to IV, as discussed in Section 7.2 of Roy and Cordy's survey [32]. While the approach is based on identification of code sequences that have a common generalisation, this is semantically aware, since it takes into account the static semantic information contained in the annotated abstract syntax tree. This makes the approach strictly stronger than Type II, but does not coincide with Type III or Type IV. In some respects our approach is *stronger* than type IV, because it supports *parametric* clones, which differ in the values of various sub-expressions that are reified into parameters of the clone function.

### 4.2 Implementation

Our implementation uses the Abstract Syntax Tree (AST) annotated with static semantic information as the internal representation of Erlang programs. Scalability, which is one of the major challenges faced by AST-based clone detection approaches, is achieved by a two-phase clone detection technique: first we use a fast string matching mechanism using generalised suffix trees to identify *clone candidates*, we then provide an accurate AST-based analysis of each candidate, ensuring that we report only true clones; that is, the system produces no false positives.

Clone detection can be applied to multi-module projects, and this process will discover not only clones within a single module but also cross-module clones with instances in different modules. More details of the implementation can be found in the original paper [22]. Wrangler also provides *incremental* clone detection [24]; the incremental version of the algorithm preserves the results of the intermediate analysis steps (ASTs, generalised suffix trees, clone tables) and reuses parts of these when they are unchanged from the previous clone detection process.

### 4.3 Modes and parameters

What does the system provide to the user who wants to detect clones in their code?

---

[2] Shown by the substitutions $\sigma_1 = \{Y \mapsto X+3, Z \mapsto 4\}$ and $\sigma_2 = \{Y \mapsto 4, Z \mapsto 5-(3*X)\}$.

**The pragmatics of clone detection and elimination**

*Modes.* First, for both identical and similar code, two modes of operation are possible:

Detection  This operation will identify all code clones (up to identity or similarity) in a module or across a project. For each clone the common generalisation for the clone is generated in the report, and can be cut and pasted into the module prior to clone elimination.

Search  This operation allows the identification of all the code that is identical or similar to a *particular selection*, so is directed rather than speculative.

Looking at examples as the system developed showed that both modes would be relevant to effective clone removal, and validation of that was provided by examples encountered in the case studies.

*Threshold parameters.* Similar code detection in Wrangler is governed by a number of threshold parameters, which describe the size and other properties of the clone.

Sequence length.  Clones identified consist of a sequence of Erlang expressions. This parameter represents the threshold value of the minimum number of expressions making up an identified clone. A value of 1 will mean all expressions (and sub-expressions) will be examined. The default value is 5.

Size of the clone.  The minimum number of tokens in the clone. This allows for small clones to be ignored. The default value is 40.

Size of the clone class.  How many times should the clone appear? The minimum is 2, which is also the default, but larger numbers can be chosen.

Number of new parameters in the clone.  This provides an upper bound for the number of places in which the clones differ: for instance in the example in Section 4.1, two parameters are introduced as the code differs on both sides of the + symbol. The default (maximum) value here is 4.

Similarity.  How large is the abstraction compared to the clone instances? We call the minimum of these ratios the *similarity*, and use a default value of 0.8 in the tool.

### 4.4 Clone reports in Wrangler

Figure 1 shows a typical clone report in Wrangler, embedded in Emacs. The upper window shows the module under examination, and the lower gives the report. For each clone we list

- Each of the instances of the clone, with a hyperlink to the instance in the upper window.
- The common generalisation, expressed as a definition of the function new_fun. The new variables in the function are named NewVar_1 etc.; note that the other variables in the function consist of the free variables from the clone, here Msg and N. which also need to be abstracted.
- For each instance, following the link we show the function call that will replace the code. This shows the function applied to the actual values of the parameters for that instance. In the boxed instance, the actual parameters are "pong!~n" and a; in the other instance (not seen) they are "ping...~n" and b.

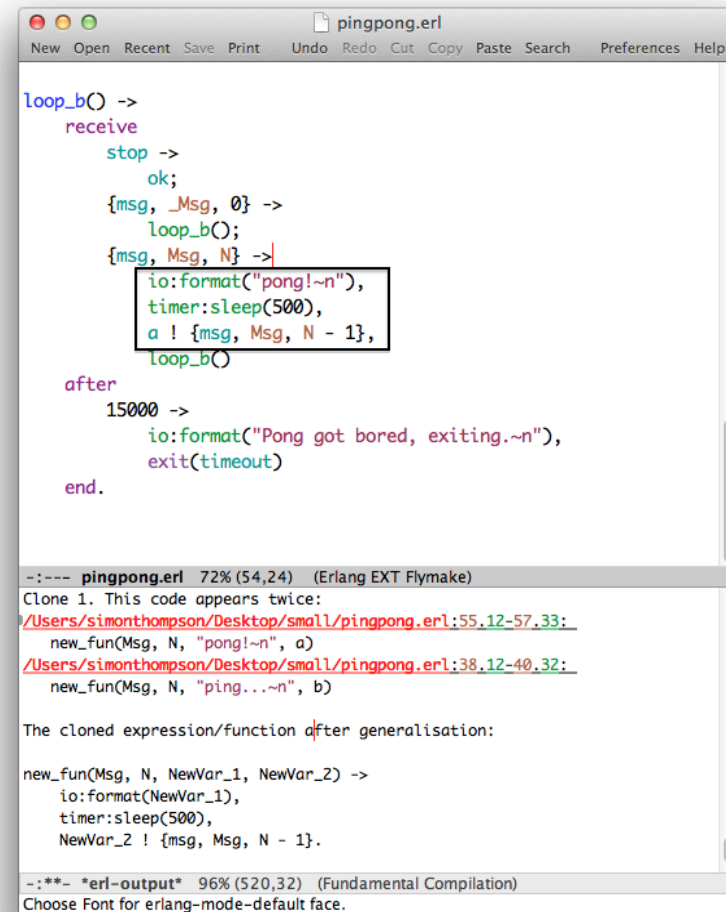

**Figure 1** Clone report in Wrangler

### 4.5 Eliminating the clone

The clone-handling functionality provided by Wrangler is in two separate components: *clone detection* results in a report that indicates clone instances, and produces a function definition that encapsulates each such instance. In the running example it is:

new_fun(Msg, N, NewVar_1, NewVar_2) ->
   io:format(NewVar_1),
   timer:sleep(500),
   NewVar_2 ! {msg, Msg, N - 1}.

Refactoring the code to *remove the clone* is provided by a series of entirely separate refactorings, after the definition has been added to the appropriate module. These are general refactorings, and there will be individual preconditions for each of them: renamings will require that no name clashes are produced, for example, and function 'folding' will also need to avoid damage to the binding structure of the program,

**The pragmatics of clone detection and elimination**

but these issues are of general concern, and nothing specifically to do with clone elimination.

In order to eliminate this clone, the definition needs to be pasted into the appropriate module, and using a series of refactorings the clone can be eliminated:

- *Rename Function* is used to rename new_fun.
- *Rename Variable* is used to rename NewVar_1, NewVar_2, etc.
- It is also possible to rearrange the variables, either by hand, or using the *Swap Arguments* refactoring [23].
- Finally, each of the instances of the clone can be 'folded' against the clone function using the *Fold Expression against Function* refactoring. This operation is performed interactively, so that the user can choose precisely which of the clone instances is to be removed. It is also possible that individual instances of this folding may fail, perhaps because of naming clashes, but these do not affect the overall correctness of the transformation since in such a case the code simply remains unchanged/

A recent extension of Wrangler for 'scripting' complex sequences of refactorings [25] allows such sequences of steps to be expressed as single refactorings within Wrangler.

## 5 Case Study I

The first case study examined part of an Erlang implementation of the Session Initiation Protocol (SIP) [34]. As a part of SIP message processing it is possible to transform messages by applying rewriting rules to messages. This SIP message manipulation (SMM) is tested by a test suite contained in the file smm_SUITE.erl, which is our subject here.

The case study was undertaken by the authors, with Li and Thompson supplying Wrangler expertise and Schumacher having knowledge about the application area.[3] The code consisted of a dozen test cases, which had been written through time by a number of authors, and it was clear that a number of these had been produced using a 'copy, paste and modify' approach.

We were looking to make the code more comprehensible and maintainable, through the introduction of appropriate abstractions into the code. This would have the consequence of making subsequent tests substantially easier to write. Clone detection was the mechanism used to identify appropriate code for extraction into functions.

The size of the sequence of versions of the files – in lines of code – is indicated in Figure 2, which shows that the code has been reduced by about 25% by means of these transformations. As we discuss at the end of this section there is still considerable scope for clone detection and elimination, and this might well reduce the code by a further few hundred lines of code.

---

[3] Adam Lindberg also contributed application area expertise.

8-**10**

**Figure 2** File sizes in case study 1.

| Version | LOC  | Version | LOC  | Version | LOC  |
|---------|------|---------|------|---------|------|
| 1       | 2658 | 6       | 2218 | 10      | 2149 |
| 2       | 2342 | 7       | 2203 | 11      | 2131 |
| 3       | 2231 | 8       | 2201 | 12      | 2097 |
| 4       | 2217 | 9       | 2183 | 13      | 2042 |
| 5       | 2216 |         |      |         |      |

**Figure 3** Initial clone data for case study 1, using the default values of the threshold parameters.

|                        | Size (LOC) | Occurrences | Total parameters | New parameters |
|------------------------|------------|-------------|------------------|----------------|
| Median                 | 17         | 2           | 4.5              | 2              |
| Mean                   | 19.1       | 3.4         | 4.8              | 2.3            |
| Maximum                | 89         | 16          | 11               | 4              |
| Minimum                | 7          | 2           | 0                | 0              |
| Largest clone          | 89         | 2           | 2                | 2              |
| Second largest         | 61         | 3           | 3                | 3              |
| Most occurring clone   | 7          | 16          | 0                | 0              |
| Second most occurring  | 9          | 14          | 1                | 1              |
| Most parameterised     | 21         | 2           | 11               | 4              |
| Number of clones       | 42         |             |                  |                |

The nature of the clones in the initial – unchanged – version of the project is shown in Figure 3, where it can be seen that the clones vary considerably in size, in frequency of occurrence and in degree of parametrisation.

## 5.1 The sequence of transformations

We now give an overview of the particular steps taken in refactoring the SMM test code, highlighting the lessons for clone detection and elimination in greyed boxes as we go along.

**Step 1**
We begin by generating a report on similar code in the module. 31 clones are detected, with the most common being cloned 15 times. The generalisation suggested in the report is shown in Figure 4 which, because it has no formal parameters, shows that

**The pragmatics of clone detection and elimination**

■ **Figure 4** Clone identified at Step 1.

```
new_fun() ->
    SetResult = ?SMM_IMPORT_FILE_BASIC(?SMM_RULESET_FILE_1, no),
    ?TRIAL(ok, SetResult),
    %% AmountOfRuleSets should correspond to the amount of rule sets in File.
    AmountOfRuleSets = ?SMM_RULESET_FILE_1_COUNT,
    ?OM_CHECK(AmountOfRuleSets, ?MP_BS, ets, info, [sbgRuleSetTable, size]),
    ?OM_CHECK(AmountOfRuleSets, ?SGC_BS, ets, info, [smmRuleSet, size]),
    AmountOfRuleSets.
```

the code is *literally* repeated sixteen times. This function definition can be cut and pasted into the test file, and all the clones folded against it.

> **Clone functions need to be named by domain experts**
>
> Of course, this new_fun needs to be renamed, and this can only be done with domain expertise: no automated tool could pick a name. In this case we choose to call it import_rule_set_file_1 since the role of this function is to import rule sets which determine the actions taken by the SMM processor, and this import is a part of the common setup in a number of different test cases. The function can be renamed using the *Rename Function* refactoring either when it is first introduced, or after it has been used to replace the clone instances.

**Step 2**
Looking again at similar code detection we find a twelve line code block that is repeated six times. This code creates two SMM filters, and returns a tuple containing names and keys for the two filters. This is a common pattern, under which the extracted clone returns a tuple of values, which are assigned to a tuple of variables on function invocation, thus:

```
{FilterKey1, FilterName1, FilterState, FilterKey2, FilterName2}
    = create_filter_12()
```

> **Choose specific names**
>
> We have named the function create_filter_12; this reflects a general policy of not trying to anticipate general names for functions when they are introduced. Rather, we choose the most specific name, changing the name to a more general one – or indeed generalising the functionality itself — at a later stage if necessary, using Wrangler.

**Step 3**
At this step a 21 line clone is detected, as shown in Figure 5; inspecting this identifies a smaller clone, encapsulated in the function shown in Figure 6, and which we choose to replace first. This was replaced by the function shown in Figure 7, where as well as renaming the function and variable names, the order of the variables is changed. This can be done simply by editing the list of arguments, because before folding against the function there are no calls to it, since it is newly introduced.

8-12

**Figure 5** Clone identified at Step 3.

```
new_fun() ->
   {FilterKey1, FilterName1, FilterState, FilterKey2, FilterName2}
      = create_filter_12(),
   ?OM_CHECK([#smmFilter{key=FilterKey1,
         filterName=FilterName1,
         filterState=FilterState,
         module=undefined}],
      ?SGC_BS, ets, lookup, [smmFilter, FilterKey1]),
   ?OM_CHECK([#smmFilter{key=FilterKey2,
         filterName=FilterName2,
         filterState=FilterState,
         module=undefined}],
      ?SGC_BS, ets, lookup, [smmFilter, FilterKey2]),
   ?OM_CHECK([#sbgFilterTable{key=FilterKey1,
          sbgFilterName=FilterName1,
          sbgFilterState=FilterState}],
      ?MP_BS, ets, lookup, [sbgFilterTable, FilterKey1]),
   ?OM_CHECK([#sbgFilterTable{key=FilterKey2,
          sbgFilterName=FilterName2,
          sbgFilterState=FilterState}],
      ?MP_BS, ets, lookup, [sbgFilterTable, FilterKey2]),
   {FilterName2, FilterKey2, FilterKey1, FilterName1, FilterState}.
```

**Figure 6** Smaller clone identified at Step 3.

```
new_fun(FilterState, FilterKey2, FilterName2) ->
   ?OM_CHECK([#sbgFilterTable{key=FilterKey2,
          sbgFilterName=FilterName2,
          sbgFilterState=FilterState}],
      ?MP_BS, ets, lookup, [sbgFilterTable, FilterKey2]).
```

**Threshold parameter choice**

Clone detection has been used with the default parameters in this case study, and the smaller clone was not identified as such. There's a trade off between lowering the parameters – thus broadening the choice of possible clones – and increasing the difficulty of identifying the clones to work on.

**Clone selection: *bottom up* or *top down*?**

As a general principle we found it more useful to replace clones bottom up, rather than top down. One reason for this is that it is easier to understand the operation of a smaller code fragment, and so to name it appropriately. Once smaller clones are replaced by appropriately named function calls it becomes easier to name larger clones.

**The pragmatics of clone detection and elimination**

■ **Figure 7**  Replacement function used at Step 3.

```
check_filter_exists_in_sbgFilterTable(FilterKey, FilterName,FilterState) ->
  ?OM_CHECK([#sbgFilterTable{key=FilterKey,
          sbgFilterName=FilterName,
          sbgFilterState=FilterState}],
      ?MP_BS, ets, lookup, [sbgFilterTable, FilterKey]).
```

> **Two modes of operation**
>
> This example shows the value of the two modes of operation of the tool. We start off by using 'detection' mode, which identifies the larger clone. This allows us to see a difference opportunity, and we can investigate this using 'search' mode that finds all clones of an identified program fragment.
>
> **Variable names and ordering need to be selected**
>
> Here we choose to rename the variables in the function identified by clone detection; this is particularly necessary when they are NewVar_1, NewVar_2 etc.

**Steps 4–5**
Introduce two variants of check_filter_exists_in_sbgFilterTable:
- In the first the check is for the filter occurring only once in the table, so that a call to ets:tab2list replaces the earlier call to ets:lookup.
- In the second the call is to a different table, sbgFilterTable being replaced by smmFilter.

> **How much to generalise?**
>
> Arguably these three alternatives could have been abstracted into a common generalisation, but it was felt by the engineers that each of the three functions encapsulated a meaningful activity, whereas a generalisation would have had an unwieldy number of parameters as well as being harder to name appropriately.

**Step 6**
Erlang provides two mechanisms for finding out whether the code for a module M is loaded:

```
erlang:module_loaded(M) -> true | false
code:is_loaded(M) -> {file, Loaded} | false
```

Use of the former is deprecated outside the code server, but both are used in this file. We want to *remove the deprecated calls*, all of which are symbolic calls in contexts like:

```
?OM_CHECK(false, ?SGC_BS, erlang, module_loaded, [FilterAtom1])
```

So what we do is to define a new function, in which we abstract over module name, the type of blade and the expected result of the call to erlang:module_loaded.

**Figure 8** New version of the code_is_loaded from Step 6.

```
code_is_loaded(BS, ModuleName, false) ->
    ?OM_CHECK(false, BS, code, is_loaded, [ModuleName]).
code_is_loaded(BS, ModuleName, true) ->
    ?OM_CHECK({file,atom_to_list(ModuleName)}, BS,
        code, is_loaded, [ModuleName]).
```

**Figure 9** Handwritten generalisation from Step 7.

```
code_is_loaded(BS, om, ModuleName, false) ->
    ?OM_CHECK(false, BS, code, is_loaded, [ModuleName]);
code_is_loaded(BS, om, ModuleName, true) ->
    ?OM_CHECK({file, atom_to_list(ModuleName)},
        BS, code, is_loaded, [ModuleName]);
code_is_loaded(BS, ch, ModuleName, false) ->
    ?CH_CHECK(false, BS, code, is_loaded, [ModuleName]);
code_is_loaded(BS, ch, ModuleName, true) ->
    ?CH_CHECK({file, atom_to_list(ModuleName)},
        BS, code, is_loaded, [ModuleName]).
```

```
code_is_loaded(ModuleName, BS, Result) ->
    ?OM_CHECK(Result, BS, erlang, module_loaded, [ModuleName]).
```

We then fold against this definition to remove all calls to erlang:module_loaded, expect for that in the definition of code_is_loaded itself. We can then write a *different* definition of this function, see Figure 8, which implements the same functionality using the other primitive. At this point it is possible to stop, having introduced the code_is_loaded function. Alternatively, in order to keep the code as close as possible to its previous version, we can *inline* this function definition. In the next step we will see another reason for doing this inlining.

**Step 7**

We note that as well as finding symbolic calls to code:is_loaded within the OM_CHECK macro call, it is also called within CH_CHECK. We are unable to replace a macro call by a variable, and so we write – by hand – a generalisation in which the macro call is determined by an atom parameter, as shown in Figure 9.

> **When to generalise?**
>
> It is here that inlining of the code_is_loaded function in step 6 is valuable: it allows us to deal with *premature generalisation*, under which we find that we want further to generalise a function without layering a number of intermediate calls: we inline the earlier generalisations and then build the more general function in a single step.

**The pragmatics of clone detection and elimination**

■ **Figure 10** Sub-clone identified in Step 8.

```
check_add_rule_set_to_filter(FilterKey, FilterName, RuleSetName,
         FilterRuleSetPosition, Result) ->
  AddResult =
    ?SMM_ADD_RULE_SET_TO_FILTER(FilterKey, FilterName,
         RuleSetName, FilterRuleSetPosition),
  ?TRIAL(Result, AddResult).
```

■ **Figure 11** Two sub-clones, as identified in Step 9.

```
check_ruleset_name_in_filter(FilterName, RuleSetName) ->
  {ok, RuleSetKey} = ?SMM_NAME_TO_KEY(sbgRuleSetTable, RuleSetName),
  check_ruleset_key_in_filter(RuleSetKey, [[FilterName]]),
  RuleSetKey.

check_ruleset_key_in_filter(RuleSetKey, Result) ->
  ?OM_CHECK(Result,
      ?MP_BS, ets, match, [sbgIsmFilterRuleSetUsageTable,
          {'_', {RuleSetKey, '_'}, '_', '$1'}]).
```

**Steps 8,9**

In these steps, a ten line clone was identified, but rather than replacing that – which combines a number of operations – it was decided to look at its sub-clones. A *sub-clone* is a subsequence (or subexpression) of an identified clone, which will *a fortiori* be a clone, potentially with a larger number of instances. Looking at this led to the identification of code used 22 times in the module, extracted as the function shown in Figure 10. This gives the ninth version of the code, and two similar sub-clones are extracted, as shown in Figure 11, giving the tenth version of the code in the study.

> **Clone choice**
>
> It is clear that we need to view the results of the clone identification in more than one way. Wrangler presents the information ordered in two ways: by the size of the clone – that is the size of the instances – and by the number of times that the clone occurs. Each criterion can be significant in different situations.

**Step 10**

Clone detection now gives the clone candidate shown in Figure 12. The body of this function sets up four rule sets and adds them to a filter, but the function identified contains extraneous functionality at the start and end:
- the filter key is created as the first action: FilterKey = …, and
- (in at least one of the clones) the rulesets are removed thus: NewVar_1 .

■ **Figure 12** The clone candidate identified in Step 10.

```
new_fun(FilterName, NewVar_1) ->
   FilterKey = ?SMM_CREATE_FILTER_CHECK(FilterName),
   %%Add rulests to filter
   RuleSetNameA = "a",
   RuleSetNameB = "b",
   RuleSetNameC = "c",
   RuleSetNameD = "d",
   ... 16 lines which handle the rules sets are elided ...
   %%Remove rulesets
   NewVar_1,
   {RuleSetNameA, RuleSetNameB, RuleSetNameC, RuleSetNameD, FilterKey}.
```

■ **Figure 13** The function extracted at Step 10.

```
add_four_rulesets_to_filter(FilterName, FilterKey) ->
   RuleSetNameA = "a",
   RuleSetNameB = "b",
   RuleSetNameC = "c",
   RuleSetNameD = "d",
   ... 16 lines which handle the rules sets are elided ...
   {RuleSetNameA, RuleSetNameB, RuleSetNameC, RuleSetNameD}.
```

and so, instead, the body is extracted as the code shown in Figure 13, in which the final action doesn't need to be passed in as a parameter, and the FilterKey becomes a parameter rather than a component of the result.

> **'Widows' and 'orphans'**
>
> Clone detection software will identify the maximal duplicated segments, but as noted in the example it may be that a maximal segment may include code that is extraneous to the core functionality identified. Again, the search mode of the tool can be used to identify the clones of the core functionality. The next step presents a second example of this phenomenon.

**Step 11**
In a similar way to the previous step, a lengthy clone is identified, but in the abstraction the first line is omitted, making it an abstraction without parameters: setup_rulesets_and_filters.

**Step 12**

This final step consists of a sequence of stages concerned with refactoring the form of

data representation in dealing with
- sets of attributes, which are transformed into named lists, such as

**The pragmatics of clone detection and elimination**

■ **Figure 14**   The function extracted at Step 12.

```
import_warning_rule_set(main) ->
  %% Since the rule set file contains errors, no rule sets will be imported.
  ...
  AttributeNames = [line_number, column, type, failing_rule_set, error_reason],
  Key1 = 1,
  ?ACTION("Do SNMP Get operations on the"
    "sbgRuleSetErrorTable with key ~p.~n", [Key1]),
  GetResult1 = ?SMM_SMF_GET(rule_set_error, [{key, Key1}], AttributeNames),
  LineW1 = 4,
  ColumnW1 = ?SMM_RULE_SET_WARN_1_COL,
  TypeAtom = warning,
  TypeMibVal = ?SMM_REPORT_TYPE(TypeAtom),
  FailingRuleSetW1 = " ",
  ResasonW1 = ?SMM_RULE_SET_WARN_1_REPORT(LineW1),
  Attributes_1 = [LineW1, ColumnW1, TypeAtom, FailingRuleSetW1, ResasonW1],
  ?TRIAL(lists:zip(AttributeNames, Attributes_1), GetResult1),
  ... 40 lines elided ... .
```

   Attributes_1=[LineW1, ColumnW1, TypeAtom, FailingRuleSetW1, ResasonW1],
- other sets of attributes are represented by 0-ary functions:
   ruleset_error_attributes() -> [?sbgRuleSetErrorLineNumber, ..., ...

Finally, we can replace an explicit zipping together of lists by the uses of function lists:zip/2, and this gives us a much better structured function shown in Figure 14.
   An incidental benefit of the inspection was to reveal that the lines

TypeAtom = warning,
TypeMibVal = ?SMM_REPORT_TYPE(TypeAtom),

were repeated, doubtless a cut-and-paste error, which went undetected because the pattern matches succeeded at the second occurrence.

> **Clone elimination and other refactorings**
>
> Clone detection as implemented in Wrangler will only find clones that consist of code segments that have a common generalisation. In practice there may be small changes that prevent larger clones being identified; it is therefore important in practice to pursue clone elimination in conjunction with other refactorings, as shown in this particular step.

## 5.2  Continuing the case study: further clone detection

The work reported here produced a sequence of twelve revisions of the code, but it is possible to make further revisions. In this section we look at a selection of reports from similar and identical code search, and comment on some of the potential clones identified.

■ **Figure 15** Clone report discussed in Section 5.2.

/Users/simonthompson/Downloads/smm_SUITE13.erl:1755.4-1761.50:
This code has been cloned twice:
/Users/simonthompson/Downloads/smm_SUITE13.erl:1772.4-1778.50:
/Users/simonthompson/Downloads/smm_SUITE13.erl:1789.4-1795.50:

The cloned expression/function after generalisation:

```
new_fun(FilterAtom1, FilterAtom2, NewVar_1, NewVar_2, NewVar_3) ->
   NewVar_1,
   code_is_loaded(?SGC_BS, om, FilterAtom1, NewVar_2),
   code_is_loaded(?MP_BS, om, FilterAtom1, false),
   code_is_loaded(?SGC_BS, om, FilterAtom2, NewVar_3),
   code_is_loaded(?MP_BS, om, FilterAtom2, false).
```

**Similar code**

The similar code detection facility with the default parameter values reports 16 further clones, each duplicating code once. The total number of duplicated lines here is 193, and so a reduction of some 145 lines could be made by replacing each clone into a function definition plus two function calls.

Looking for similar code with the similarity parameter reduced to 0.5 rather than the default of 0.8 reports 47 clones, almost three times as many. Of these, eight duplicate the code twice (that is, there are *three* instances of the code clone) and some of these provide potential clients for replacement. However, not all of them appear to be good candidates for replacement. For example, consider the report shown in Figure 15. This code has three parameters, and the first is an arbitrary expression, NewVar_1, evaluated first in the function body. A more appropriate candidate is given by omitting this expression, and giving the generalisation

```
new_fun(FilterAtom1, FilterAtom2, NewVar_1, NewVar_2) ->
   code_is_loaded(?SGC_BS, om, FilterAtom1, NewVar_1),
   code_is_loaded(?MP_BS, om, FilterAtom1, false),
   code_is_loaded(?SGC_BS, om, FilterAtom2, NewVar_2),
   code_is_loaded(?MP_BS, om, FilterAtom2, false).
```

This particular generalisation has a similarity score of more than 0.8, but does not appear in the default report because it involves only 4 expressions, and the default cut-off is a sequence of at least 5.

### 5.2.1 Identical code

The standard report to detect identical code reports 87; the number is larger here because the default threshold for reporting here is to consist of at least 20 tokens rather than 5 expressions or more. However, a number of *over-generalisations* result from this report; for example, the function

```
new_fun(ModuleName1, NewVar_1, NewVar_2, NewVar_3, NewVar_4) ->
```

**The pragmatics of clone detection and elimination**

```
code_is_loaded(?SGC_BS, NewVar_1, ModuleName1, NewVar_2),
code_is_loaded(?SGC_BS, NewVar_3, ModuleName1, NewVar_4).
```

is reported as occurring 23 times. Arguably, generalising to replace these two expressions will not result in code that is more readable than it is already: it clearly states that it checks for two pieces of code being loaded.

**5.2.2 Carrying on**

There are clearly some more clones that might be detected, but as the work progresses the effort involved in identifying and replacing code clones becomes more than the value of transforming the code in this way, and so the engineers performing the refactoring will need to decide when it is time to put the work aside.

# 6 Lessons learned

This section highlights the lessons learned during the activity reported in the previous section, cross-referring to the steps of that process when appropriate.

**6.1 Inlining is a useful refactoring**

There is a clear use case for function *inlining* or *unfolding* when performing a series of refactorings based on clone elimination [Step 7]. The scenario is one of *premature generalisation* thus:

- identify common code, and generalise this, introducing a function for this generalisation;
- subsequently identify that there is a further generalisation of the original code, which could benefit from being generalised;
- the problem is that some of the original code disappeared in the first stage of generalisation, and so it needs to be inlined in order to generalise it further.

Of course, it would be possible to keep the intermediate generalisation as well as the final one, but in general that makes for less readable code, requiring the reader to understand two function definitions and interfaces rather than one.

Inlining is also useful to support a limited form of API migration [Step 6].

**6.2 Bottom-up is better than top-down**

We looked at ways in which clones might be removed, and two approaches seem appropriate: bottom-up and top-down. In the latter case we would remove the largest clones first, while in the other approach we would look for small clones first, particularly those which are the most common. We decided to use the *bottom-up* approach for two reasons [Steps 3, 8, 9].

- Using this it is much easier to identify pieces of functionality which can easily be *named* because they have an identifiable purpose.

■ **Figure 16** An example of self-documenting code.

BlahMeaning = ... blah ... ,
FooMeaning  = ... foo ... ,
Result = f(BlahMeaning, FooMeaning)

- it is also likely that these will not have a huge number of parameters, and in general we look for code which is not over-general.

Finally, there is the argument that – to a large extent, at least – it should not matter about the order in which clone removal takes place, since a large clone will remain after small clones are removed, and *vice versa*.

### 6.3 Clone removal cannot be fully automated

What we have achieved in this example is clearly *semi-automated*: we have the Wrangler support for identifying candidates for clones but they may well need further analysis and insight from uses to identify what should be done. For example,

- Is there a spurious last action which belongs to the next part of the code, but which just happens to follow the clone when it is used? If so, it should not be included in the clone [Steps 10,11]. This can also apply to actions at the start of the identified code segment.
- Another related reason for this might be that this last operation adds another component to the return tuple of an extracted function; we should aim to keep these small.
- We might find that we identify behaviour of interest which occurs close to an identified clone; then use similar expression search to explore further.
- An identified clone may contain two pieces of separate functionality which are used together in many cases, but not in all cases of interest. Because of the thresholding of clone parameters it might be that we only see the larger clone because the smaller one is below the threshold chosen for the similar code report [Steps 3, 8, 9].

### 6.4 Self-documenting code

Rather than using the telegraphic

Result = f( ... blah ... , ... foo ... )

it is always useful to name values, as shown in Figure 16, since the latter case is *self documenting* in a way that the former is not. On the other hand, is it the responsibility of the client of an API, here for the function f, to document this API at its calling points? If it is to be documented at a calling point it can be done by choice of variable names, or by suitably placed comments.

**6.5 Naming values**

How should constants best be named in Erlang code. Three options are possible: namely definition by

- a macro definition
- a function of arity 0
- a local variable

The code at hand does the first and last: the second was added during the refactoring process.

Another option presents itself: instead of worrying about what is contained in a set of variables, should a record be used instead? This has some advantages, but records have drawbacks in Erlang. Names are not first class, so cannot be passed as parameters or values, which is something that can be used in practice, such as passing a list of field names to the zip/2 function [Step 13].

**6.6 Improvements to Wrangler**

The case study also brought to light a number of improvements that might be made to Wrangler. Inlining was the most important, and is now included in the latest release of the system.

It was also suggested that a number of options could be added to the *Code Inspector*: this highlights "bad smells" and other notable code features, such as: variables used only once, variables not used, and rebinding of variables (that is, bound variables occurring in a pattern match).

Finally, the question was raised about how much refactoring sequences used on one module could be *reused* in refactoring another. This question of memoisation or scripting merits further work.

# 7 Improvements added subsequently

Visits to the product development group at Ericsson in Linköping in 2010 allowed us to explore duplicate code detection and elimination in a variety of test suites. We were also able to use function extraction to add structure to long functions by breaking them into a sequence of calls to smaller (extracted) functions.

These visits coincided with the development of the module structure exploration tools in Wrangler [21]; we were able to explore inter-module dependencies and in particular to examine circular dependencies. We noted that some cycles in the module dependency structure are legitimate; an example would be two ends of a message interaction that are implemented in different modules.

On the basis of the work done during these visits, a number of improvements and bug-fixes were added to Wrangler.

- Add the facility to introduce a local definition, and to inline instances of a local definition.

- In reporting on clones, as well as presenting the new function that represents the least common generalisation, the report should also include the actual parameter values for each instance.
- Provided a full set of key bindings for all the refactorings in Emacs. This follows the scheme C-c C-w X, with C-c C-w as the 'Wrangler command' common prefix.
- Clone detection is controlled by a minimum number of expressions in the (cloned) expression sequence. This can miss the case of a single-expression clone, and so we have also introduced a minimum number of tokens for clone size which can be used in conjunction with clone search over sequences of any length.
- Introduce another parameter for clone detection: the number of parameters to the common abstraction. There are two choices for this: it could either be the maximum number of parameters for the clone, or the number of *new* variables introduced by the generalisation, excluding those variables that are already unbound in the clone instances.
- Handle side-effects in clone extraction: this enhancement ensures that side effects are handled properly in clone extraction by wrapping up in a closure any actual parameter which can have a side effect.
- In clone detection it is possible that when two variables have the same values in each instance, and so can be replaced by a single variable: this has been implemented.
- In clone detection across multiple modules it is necessary to check the use of function names: it is not always the case that f/N in two modules means the same thing; in fact, it most usually doesn't (except for BIFs). On the other hand, it is possible that m:f/N and f/N are the same (precisely when f/N is used in module m). This name resolution is now correctly implemented.
- The order of the variables in the return tuple of a generalisation function was modified so that it reflected the order in which they are declared in the function body.
- Users of version control systems requested reporting of when a file is not writeable, as this will probably invalidate a refactoring step.

Other suggestions were discussed; some were scheduled for future implementation, and others were rejected.

- Refactoring to move definitions between scopes: that is from local to top-level and *vice versa*. In a language like Haskell with a more developed facilities for scope nesting this is a necessity, but in Erlang we see it as being of less practical value.
- On the basis of analysis of forward and backward dependencies within expression sequences, it is possible to provide refactorings to move statements "up" and "down" in an expression sequence. One use case is to do this to assist with clone detection, when an inserted statement splits a larger clone in two. An example of this is the subject of Section 8.1, where it is evident that cases like this require manual intervention when they are extracted in practice.
- Definitions of records and macros in Erlang can be contained in Erlang source files, or included in header files. Providing the facility to detect cloned macro or record

- definitions across multiple modules could be complemented by the facility to move these into a common header file. This is future work.
- Source files contain comments – some of which are 'semi-structured' information used in the edoc documentation system – and also type specifications (-spec statements). While it makes little sense to try to refactor free text comments, it could be useful to indicate to users which comments might have been invalidated by a refactoring.
  With the advent of release R14B of the Erlang system there is now a single spec notation for both specifications and comments, and these are to be transformed as a part of refactorings from a future Wrangler release.
- In building the common abstraction for a clone, we choose to name variables as `Newvar_N`: could we do better? If we know it's a module or function name, then that is possible; otherwise we could take a name from one of the instances, if any instance is itself a variable. We decided, however, that this is potentially misleading, and that the `Newvar_N` names force the user to choose names that accurately reflect the semantics of the variable.
- A report on variable use within function bodies could be used when it is desirable to split – by means of function extraction – function definitions that are too long. We plan to investigate this.

## 8 Following up: continuing the case study

The original visit by the Kent team to Lindberg and Schumacher at Ericsson in Stockholm in 2009 was followed up by visits to Ericsson Linköping in 2010 and 2011, where again the Wrangler team (Thompson and Li) sat together with developers to restructure their code.

This allowed them to replicate the study reported in Section 5, to gain feedback on improvements of Wrangler implemented after the initial case study, and, finally, to understand the main obstacles to wider adoption of Wrangler within Ericsson and more widely.

### 8.1 Clone detection

#### 8.1.1 Multi-module search.
Our initial experiment was to look for clones on a set of code of more than 1MB across some dozen files. Generating the clone detection report took approximately 20 minutes: we found 1066 clone candidates (through the initial string-based search) of which 457 were found to be true clones (by analysis of the appropriate parts of the annotated abstract syntax trees).

The format of the clone report, which orders the results on the size of the clone set and on the size of the clone members, means that the report cannot be shown until *all* clones are detected. It was frustrating to watch the clone checks tick past while being unable to see any of the results until all were complete.

Wrangler is being updated to stream the clone results to standard output as they are produced, so that the system remains responsive while processing. The full report will be written to a file, formatted and ordered, once all the results are processed. In the context of a multi-module clone search, the report is also to be upgraded to single out the inter-module clones from the intro-module clones.

### 8.1.2  Single module search: overview.

We then looked at clones within one of the modules consisting of some 2500 lines of code. 74 initial candidates identified, of which 14 were true clones.

In examining clones it was very helpful to use the "Instances of a variable" functionality in the Wrangler inspector menu. This allows programmers to view all the uses of a variable, distinguishing instances of the variable from the defining occurrence(s). Together with the clone report this shows where the clone instances differ, and what the instantiations of the newly introduced variables are. This only works on the new definition once it is pasted into the module, but would be helpful as a part of the report too, so that this information can be used in assessing the clone for refactoring.

### 8.1.3  Single module search: clone choice.

We were presented with a choice of fourteen clones. We began by looking at the largest clone classes, and rejected the first candidate (occurs 7 times) but instead opted for the second. This was the choice of the developers to whom it seemed to be a more complete piece of functionality; they also noted that "one of the most difficult things [in the clone extraction process] is naming the parameters".

### 8.1.4  Growing the largest clone.

The largest clone was some 100 lines of code which appeared twice. These formed the latter parts of two tests, so that the code in both cases went right to the end of the test. This focussed attention on how the tests differed, and with a combination of Erlang and domain expertise the team was able to refactor the tests by hand so that their entire bodies were clones. The manual refactorings were these:

- One had an added line: this turned out to be setting an attribute of the system under test (SUT) and occurred next to another attribute setting; in the code these were manually changed to calls which set a *list* of attributes in a single call. This function was already available in the API of the SUT, but could have been added if necessary.
- The result of a function call was tested by a pattern match of the form
    {V,result_list,X}
  
  with a different result_list in the two cases. Wrangler didn't spot this as a clone - it is, in fact, because it is possible to match against a bound variable in a pattern in Erlang. To work around this we introduced local variable definitions of the result_list values and then performed a pattern match against a tuple of variables in each of the tests.

**The pragmatics of clone detection and elimination**

■ **Figure 17** Clone instances as function calls.

```
new_fun(Config,
   [{cellTraceFileSize, 100}],
   ["INTERNAL_TESTEVENT_UE", "INTERNAL_TESTEVENT_EXT"], 21,
   "INTERNAL_EVENT_MAX_FILESIZE_ROVERY")

new_fun(Config,
   [{totalCellTraceStorageSize, 100},
    {cellTraceFileSize, 100}],
   ["INTERNAL_TESTEVENT_EXT"], 50,
   "INTERNAL_EVENT_MAX_FILESIZE_RECOVERY")
```

### 8.1.5 Naming the clone and its parameters.

This results in a generalisation with five variables: one which is a parameter of each of the cloned tests, and the other four indicating differences between the two. The calls are shown in Figure 17 and the function and variables are named as indicated in the function header for the cloned code:

```
check_cell_trace_size(Config, Attributes, Events, NrOfRuns,
    FileSizeEvent) -> ...
```

### 8.1.6 Final observations.

The clone detection process also has the effect of pointing out precisely how the clones differ: in one case an actual parameter is

```
"INTERNAL_EVENT_MAX_FILESIZE_RECOVERY"
```

while in the other it is

```
"INTERNAL_EVENT_MAX_FILESIZE_ROVERY"
```

It seems pretty clear that this latter is a corruption of the former, and that in fact they should be the same. This would mean that the parameter FileSizeEvent would be eliminated, since its actual values would coincide in the two instances.

A clone which occurs in the middle of a function definition can potentially have a large number of parameters and also return a tuple containing a large number of fields. The former is because the code in the clone can depend on variable bindings occurring in code before the clone begins, and the latter because the cloned code itself sets up bindings which are used in the remainder of the function. The effect of a clone coinciding with the entire function body is to reduce the number of parameters, and also to reduce the return value to that of the original function. This effect (specifically the effect of the number of parameters reducing) was evident as we "grew" the clone reported here.

## 8.2 Macro elimination

There is a disadvantage in using Erlang macros, in that they cannot be used in expressions evaluated in the shell. While it would be possible to support macros in the shell, in the absence of that to get the exploratory nature of shell evaluation, we needed to find some way of eliminating macro definitions.

The following mechanism was investigated, assuming the macro to be eliminated is foo(X,Y) with definition.

-define(foo(X,Y),E).

*Step 1.* Build a function definition thus

fun_foo(X,Y) -> ?foo(X,Y).

*Step 2.* Fold against the function fun_foo, turning all macro invocations into calls to fun_foo.

*Step 3.* Redefine fun_foo thus:

fun_foo(X,Y) -> E. % i.e. the RHS of the definition of the macro foo.

This approach works fine within a single module, except for 0 arity macros which will be replaced by calls with a trailing pair of parentheses, as in fun_foo().

What if the macros are in a header file? We will need to put the definition of fun_foo in a file. In general we will put it in a new file fun_macros. How is it used? ?foo(X,Y) becomes

    fun_macros:fun_foo(X,Y)

    fun_foo(X,Y) -> fun_macros:fun_foo(X,Y).

## 8.3 Obstacles to adoption of Wrangler

We see two sets of obstacles in the way of adoption of refactoring more widely. The first is a perceived lack of time to perform large-scale restructuring of systems, despite a clear need being expressed by developers, because it made their code both more readable and more maintainable by introducing structure to what was previously straight-line code. Where Wrangler has been adopted successfully it has become part of a programmer's usual workflow, rather than a separate process: this works well with incremental improvements, but larger-scale transformations are less easily integrated in this way.

A second set of obstacles come from the integration of Wrangler into the wider-scale project workflow. For example, we have been able to adapt Wrangler to work with particular source code repositories, but some systems bring particular challenges.

A third potential problem comes from the efficiency of clone detection over large code bases. To counter this we have developed an incremental clone detection approach [24]. In studies of a series of system evolution steps this approach shows a reduction of clone re-detection times to 10–20% of the original time.

## 9 Property extraction through refactoring

One of the principal goals of the ProTest project [30] is to find mechanisms that support the transition from traditional unit testing to property-based testing. One mechanism (briefly reported in an earlier paper [26]) is to use the clone detection facilities of Wrangler to support the discovery of properties within existing test suites. Not only does this provide a refactoring of unit tests into properties, but it can also indicate how the property can be generalised.

This generalisation will give a more complete description of the intended behaviour of the system, and this property can in turn be tested using randomly generated values in QuickCheck [2]. Not only does this turn tests into properties, but it will also have the effect of generating new tests for the system from the existing test suite. After describing the process in more detail we give an example of property extraction in practice.

### 9.1 Property extraction

The Wrangler approach to combining test cases into QuickCheck properties follows these steps:

- Apply Wrangler's similar code detection functionality to the test suite. The clone detector reports clone classes, each a set of code fragments in which any two of them are similar to each other. For each clone class, the clone detector also reports the least-general common abstraction of its clone instances, in the format of a new function.
- Identify a clone class which consists of complete test cases, copy and paste the least-general common abstraction function of that clone class into the test suite module, then *rename* the function, and apply the *fold expression against function* refactoring to this function. This refactoring replaces instances of a function body by a call to that function: in this case the newly introduced function.
- Apply Wrangler's *test case to property* refactoring to the least-general common abstraction function. This refactoring searches for all the application instances of this function, and collects the actual parameters. The actual parameters collected are then analysed for dependency between parameters, and transformed into QuickCheck data generators; the function itself is wrapped up, if necessary, as an assertion. A QuickCheck property is then generated by combining the data generator and the property using the QuickCheck macro ?FORALL.

### 9.2 Example

The example in an earlier paper [26] shows how a property is extracted from a set of tests identified as clones. In Section 8.1 we reported a similar clone extraction exercise, where some intervention was required to make the maximal clone identifiable, a typical experience.

■ **Figure 18** Two example sets of parameter values.

[{cellTraceFileSize, 100}]
["INTERNAL_TESTEVENT_UE", "INTERNAL_TESTEVENT_EXT"]
21

[{totalCellTraceStorageSize, 100}, {cellTraceFileSize, 100}]
["INTERNAL_TESTEVENT_EXT"]
50

■ **Figure 19** Generators for parameter values.

oneof([[{cellTraceFileSize, 100}],
    [{totalCellTraceStorageSize, 100}, {cellTraceFileSize, 100}]])
oneof([["INTERNAL_TESTEVENT_UE", "INTERNAL_TESTEVENT_EXT"],
    ["INTERNAL_TESTEVENT_EXT"]])
oneof([21,50])

■ **Figure 20** Generalised generators from Figure 19.

oneof([[{cellTraceFileSize, nat()}],
    [{totalCellTraceStorageSize, nat()},
     {cellTraceFileSize, nat()}]])
oneof([["INTERNAL_TESTEVENT_UE", "INTERNAL_TESTEVENT_EXT"],
    ["INTERNAL_TESTEVENT_EXT"]])
nat()

The calls to the clone function shown in Figure 17 give two possibilities for the parameter values, as shown in Figure 18, where we ignore the first parameter (the variable Config) and the last (the string and its corrupted version).

These choices can be generalised to the generators shown in Figure 19, and replacing the integers by arbitrary natural numbers, nat(), by those in Figure 20. This configuration can be used to generate random tests from a set including the original two, but widening this substantially. In all probability, even if the SUT is correct, some of these will pass and others will fail, since the generalisation may include illegitimate combinations of parameters; on the other hand this failure can be a result of the implementation missing "corner cases" that are plainly not covered by the two existing tests which cover typical cases of normal behaviour. In this way the generalisation can lead to a greater understanding of the behaviour of the system.

## 10  Conclusions and Future Work

As we have reported, the exercise here was only possible as a collaboration between the developers of Wrangler and engineers engaged in developing and testing the target system. Together it was possible substantially to re-engineer the test code to make it more compact and more structured. As well as illustrating the way in which Wrangler can be used, we were able to provide guidelines on refactoring test code in Erlang which can also be applied to systems written in other languages and paradigms. Pulling together the observations of the paper, we are able to make a number of conclusions.

First, it is impossible completely to automate the process of clone identification and removal. The notion of what is and isn't a clone is dependent on the choice of a set of parameters, and changing these will give a different, but related, set of clones. Even with a single choice of parameters, the identification of the clone function is subject to choice: are extraneous fragments ("widows and orphans") attached? should a sub-clone be prioritised over an enclosing one? how much generalisation take place?

Once the clone function is chosen, there is the matter of naming the function and its parameters in an appropriate way. In a number of cases it was necessary for the engineer to make some changes to the codebase in order to optimise the clones identified. None of these matters is routinely subject to automation.

We have also seen incidental effects of clone detection. It can be used as a way of comprehending an existing codebase, and also indicate errors – intentional or otherwise – in that code.

While we contend that the results are of general applicability, since the clone detection algorithm and the programming language are representative of approaches in general, for the future it would be of merit to explore other types of programming domain, to include programs written in different languages and using different clone identification mechanisms. It would also be of interest to explore larger datasets to explore the possibility of using machine learning to choose threshold values for clone identification, perhaps relative to different application domains and clone models.

**Acknowledgements**   We gratefully acknowledge the support of the 7th Framework Programme of the European Commission for the ProTest [30] (215868) and RE-LEASE [31] (287510) projects which supported the work reported here. We also wish to acknowledge the support of staff from Ericsson AB in Stockholm and Linköping, in particularly Adam Lindberg, for working with us on the case studies.

**The pragmatics of clone detection and elimination**

**About the authors**

**Simon Thompson** is Professor of Logic and Computation in the School of Computing at the University of Kent, UK. His research interests include computational logic, functional programming, testing and diagrammatic reasoning. His recent research has concentrated on all aspects of refactoring for functional programming, including the tools HaRe and Wrangler for Haskell and Erlang. He is also the author of standard texts on Haskell, Erlang, Miranda and constructive type theory. He is a Fellow of the British Computer Society and degrees in mathematics from Cambridge (MA) and Oxford (DPhil). He can be contacted at S.J.Thompson@kent.ac.uk

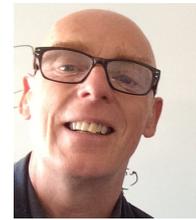

**Huiqing Li** is a visiting researcher at the school of the Computing, University of Kent. She received her B.A and Master degrees in computer science from Southeast University, China, in 1992 and 1995 respectively, and her Ph.D. degree in computer science from University of Kent in 2006. Her research interests include program refactoring, clone detection, multi-core and functional programming.

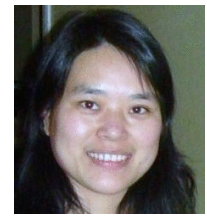

**Andreas Schumacher** is a senior specialist in platform architectures, and product owner of Erlang/OTP at Ericsson, Stockholm, Sweden. He has been working in different roles with the development of carrier-grade communication systems for over 15 years. His work on ProTest took place while he was system manager for the SIP Message Manipulation feature and a member of Ericsson Software Research. His interests include architecture principles and design of large-scale, fault-tolerant distributed systems, and the theory of concurrent and distributed systems. Andreas holds a degree in computer science and software engineering from the University of Applied Sciences, Hamburg, Germany. He is a member of IEEE Computer Society, ACM, and German Informatics Society.

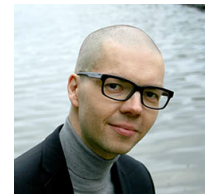